\documentclass[cameraready]{Interspeech}
\usepackage{multicol}
\usepackage{multirow}
\usepackage{dirtytalk}
\usepackage{float}

\title{No-Shot Text-to-Speech: Limitations of Zero-Shot TTS and its Evaluation Methods in Representing Queer and Transgender Voices}

\author[affiliation={1}, orcid=0009-0005-3693-511X]{Juliana}{Francis}
\author[affiliation={2}, orcid=0009-0006-4857-9498]{Robin}{Netzorg}
\author[affiliation={1}, orcid=0000-0002-0397-6442]{Joakim}{Gustafson}
\author[affiliation={1}, orcid=0000-0003-1175-840X]{Éva}{Székely}

\address{
    $^1$ KTH Royal Institute of Technology, Sweden \\
    $^2$ University of California, Berkeley, USA 
}

\email{jfrancis@kth.se, robin\_netzorg@berkeley.edu, jkgu@kth.se, szekely@kth.se}


\usepackage{comment}

\begin{document}

\maketitle
\begin{abstract}
    Gender expansive voices remain underrepresented in speech datasets, which could cause challenges with generalizability in zero-shot Text-To-Speech (TTS) systems. We evaluate six zero-shot TTS models on two datasets: one containing non gender expansive voices (N-GE) and one with gender expansive (GE) voices from self-identified speakers. Human ratings found that 2 of the 6 models synthesize less similar GE voices compared to N-GE voices, and 3 have improved performance for GE voices. Automated evaluations yield conflicting results, with speaker similarity and automated mean opinion score predictors rating performance of GE voices inconsistently between models and human ratings. This highlights a potential blind spot in evaluation pipelines, emphasizing a need for more representative data and metrics that better capture human perceptual judgments. Our findings stress the need to include transgender, queer, and minority communities in creation of better datasets and model training.
\end{abstract}

\let\thefootnote\relax\footnote{Proceedings of Interspeech 2026, 10.21437/Interspeech.2026-709}

\section{Introduction}
Artificial intelligence technologies in text-to-speech (TTS) systems and natural language processing (NLP) have enhanced human-computer interaction, often at the expense of minority groups \cite{hussain2025bias, hagerty2019global}. While these offer benefits, they inherit and amplify societal biases, leading to ethical and practical concerns that remain unaddressed \cite{mcara2024ai}. Recent studies have found that end-to-end TTS models can render `gay voices' to sound less `gay', as determined by queer people \cite{campweshould}, erasing key aspects of a speaker's identity. Research indicates that models trained on gender-specific datasets under-perform when compared to non gender-specific datasets when used for model training \cite{boito2022study}, suggesting that current data collection practices may inadvertently limit voice diversity. We evaluate zero-shot TTS performance in replicating voices from underrepresented groups, and examine whether similar biases are present in evaluation algorithms. 'Zero-shot TTS' refers to TTS systems that can generate speech for unseen speakers using a few second voice sample without additional fine-tuning. In this study we focus on “Gender-Expansive” (GE) voices, which refers to voices from speakers who self-identify as not being cisgender We use the term “Non Gender-Expansive” (N-GE) to refer to cisgender speakers. 

\subsection{Data Bias and the Necessity for Diverse Training Sets}
The ability of current state-of-the-art (SOTA) zero-shot TTS to clone voices with high fidelity is dependent upon the diversity and quality of their training data. Studies such as the Globe corpus work \cite{wang2024globe} demonstrate that models trained on diverse datasets are better able to generalize to unseen voices. Similarly, a multilingual model has shown promising results in adapting to a variety of linguistic features across different languages \cite{casanova2024xtts}. However, pervasive issues in data collection, such as sampling skew and aggregation distortions compromise the representativeness of these datasets \cite{mehrabi2021survey}. For example, the Common Voice dataset \cite{ardila2019common} reports near zero-percent participation from transgender individuals. Moreover, as an example within NLP, a large text dataset disproportionately removed queer and trans-related text, leading to data on queer people being defined by those without lived experience \cite{zhouqueer}. A barrier in developing equitable TTS and speech analysis systems is ensuring training datasets accurately represent real world voice diversity, as inadequate representation can lead to significant performance disparities \cite{wang2024globe}. Racial disparities have been found in automated speech recognition, as well as gender and dialect biases in YouTube's automatic captions \cite{koenecke2020racial, tatman2017gender}. Although transgender, queer, and other underrepresented groups constitute a smaller proportion of the population \cite{twenge2024increases, ipsos_2021_lgbt}, oversampling or other corrective measures can be essential to ensure these voices are accurately represented, but these techniques are rarely used \cite{fenu2020exploring, dheram2022toward}.

\subsection{Failure of Speech-based Machine Learning Systems}
The challenges associated with biased voice datasets extend beyond TTS systems. Deepfake audio detection algorithms often fail when confronted with out-of-distribution data \cite{muller2024harder}, which can result in cloned voices from marginalized groups being undetected, increasing risks of impersonation. In a similar vein, speaker verification systems have exhibited significant performance degradation, failing to categorize transgender individuals as the same person when they employ gender-affirming voice modification techniques \cite{netzorg2024speech}. This undermines the technical reliability of these systems and contribute to a broader sense of exclusion among affected users, reinforcing the perception that such technologies are not designed for all demographic groups \cite{feng2023review}.

\subsection{Inherited Societal Norms and the Impact on Speech Generating Device Users}
Issues of gender-induced bias in voice synthesis is especially impactful on frequent users of devices that utilize this technology, especially as voice models tend to inherit prevailing societal gender norms \cite{yeon2023gender}. When not tackled, these biases may be amplified, prompting participants in a recent study to advocate for the use of gender-neutral voices as a mitigation strategy \cite{yeon2023gender}. This issue also has a particular effect on users of speech generating devices, who already face stigmatization within society \cite{mertl2018quality}. For these individuals, having personalized voices that accurately reflect their identity is often a necessity for effective and dignified communication \cite{castellano2019, szekely2024inclusive}. 

\subsection{Research Objectives and Contributions}
Motivated by these challenges, this study addresses two key research questions:
\begin{enumerate}
 \item \textbf{Voice Cloning Bias:} Do zero-shot TTS systems have performance differences, as judged by human listeners, on GE voices?
 \item \textbf{Comparative Bias Analysis:} Does a bias exist in machine learning evaluation algorithms used for speech analysis, when compared against other evaluation metrics and subjective perceptual metrics?
\end{enumerate}

\section{Methods}

\subsection{Voice Synthesis and Preprocessing}

Six zero-shot TTS models were used to assess potential biases in synthesizing GE voices: CosyVoice2, E2TTS, F5TTS, XTTS, Zonos, and LinaSpeech \cite{casanova2024xtts, du2024cosyvoice, eskimez2024e2, chen2024f5, zyphra_2025_github, lemerle2024lina}. CosyVoice2, E2TTS, F5TTS, and XTTS were all chosen as examples of models trained on large multilingual, and potentially more diverse, datasets. ZONOS was chosen as a current industry model, and LinaSpeech is used as an example of a smaller zero-shot TTS model.

Two datasets were utilized for assessing performance. N-GE voices, the Globe Dataset \cite{wang2024globe} was used, containing assumed majority N-GE speaker samples. To select speakers from this dataset, an even split was created between \say{male} and \say{female} speakers, and only selected participants with accents labeled as \say{United States English}. Gender and accent data were provided by the metadata of this dataset. We note that the creators of the GLOBE dataset have assumed the gender of speakers who provided no gender data and assigned a gender label using a classifier model \cite{wang2024globe}. This could cause incorrect classification as these often fail for speakers whose voices fall within ambiguous ranges \cite{lebourdais2022overlapped}. For GE speakers, the `Mid-Atlantic Gender Expansive Speech' (MAGES) dataset \cite{hope_2023_midatlantic} was used which contains GE voices based on an individual's gender self-identification, and these labels are maintained in our analyses. 

Audio used as reference for generation is created by concatenating audio files from each speaker, to reach an approximate duration of 10 seconds. A reference sample generated in this way was created for all 14 MAGES speakers and 14 GLOBE speakers (chosen using the above criteria), which were then used for synthesis. This sample size is used due to the limited availability of datasets explicitly containing GE speakers.

For each speaker, 15 phrases from the Harvard sentences were synthesized using each model. All audio samples were standardized to 16 kHz and -20 dB to ensure consistency in evaluations.

\subsection{Human Evaluation}
16 participants were recruited to rate the synthesized voices. Each participant listened to a 3-4 second sample from each speaker used in the concatenated reference sample, and then asked to rate the similarity of synthesized samples to this. Subjective evaluations were conducted using a MUSHRA-like paradigm to compare across models \cite{schoeffler2018webmushra}. This MUSHRA-like implementation is a perceptual study design which presents participants simultaneously with a reference audio and a synthesized sample from each model (6 samples total) per trial, with 3 trials total per reference speaker. The order of trials was randomized. Participants rate each sample's similarity to the reference audio on a scale between 0-100. A total of 8 speakers, 4 from each dataset, were used for the MUSHRA-Like assesment, giving a total of 24 trials for a total of 144 ratings. 

\subsection{Machine Learning/Automated Metrics}
All synthesized samples were also assessed using three measures, automated mean opinion score (AMOS), speaker similarity, and word error rate (WER) \cite{cooper2022generalization, ECAPA, radford2023robust}.
speaker similarity was calculated by computing cosine similarity between the reference audio's speaker embedding and each of the synthesized sample's embedding from this reference audio. Speaker embeddings were extracted from the reference and synthesized voices using ECAPA-TDNN, TitaNet-L and ReDimNet-M (vb2+vox2+cnc-ft-mix), three models which provide speaker embeddings \cite{ECAPA,speechbrain,nagrani2017voxceleb, koluguri2022titanet, yakovlev2024reshape}.
AMOS was found using models trained to predict human mean opinion scores (MOS). For this we used UTMOS \cite{saeki2022utmos} and a finetuned wav2vec2 model for mean opinion score (MOS) prediction as described in Cooper, et al. 2022 \cite{cooper2022generalization}.
Between the speaker similarity ratings and AMOS models respectively we calculated the Intraclass Correlation Coefficient (ICC) to assess the consistency of the predicted ratings.
WER was calculated using error between the actual generated audio's transcription and the sentences used for synthesis. Transcripts were generated using the ASR system Whisper \cite{radford2023robust}.


\section{Results}

\subsection{Human Listening Results}
\begin{figure*}
 \centering
 \includegraphics[width=\linewidth]{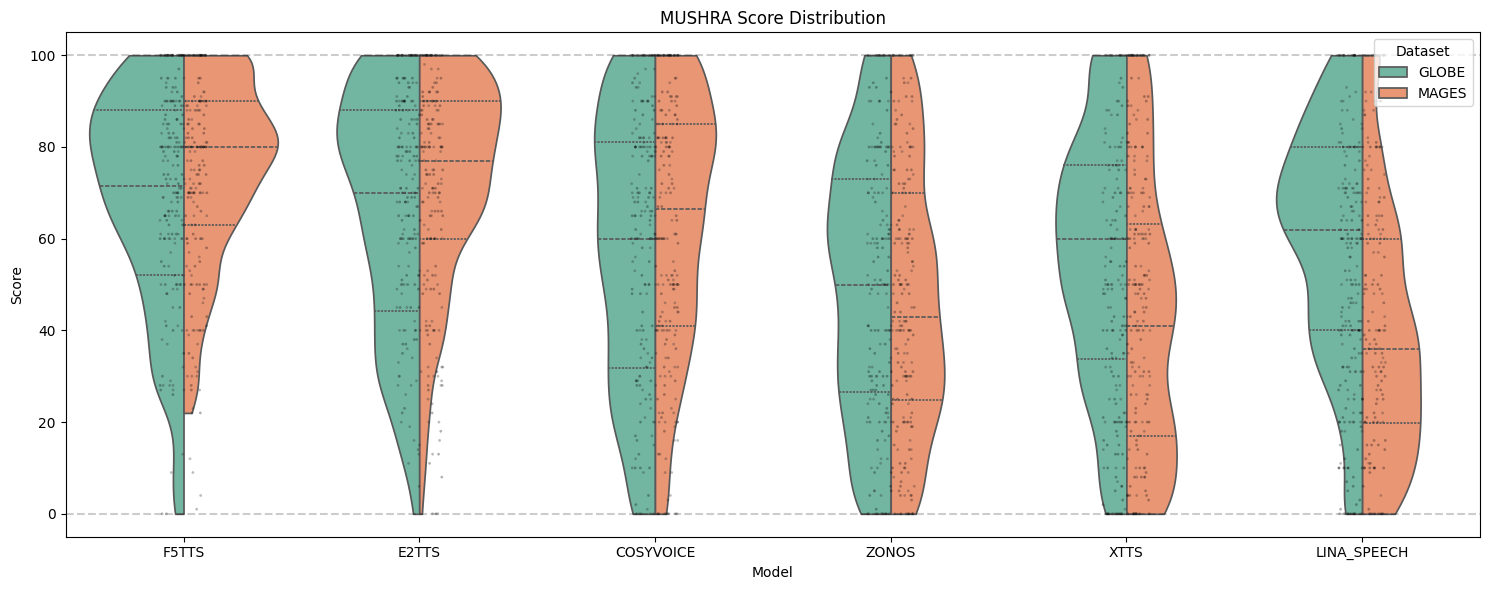}
 \caption{Human ratings for each model, split by dataset}
 \label{fig:mushra_data}
 
\end{figure*}

Results from the  listening tests are presented in Table~\ref{tab:mushra_results}. Synthesized GE voices were rated as significantly more similar to the reference audio compared to N-GE voices for E2TTS, F5TTS, and CosyVoice2; and GE voices were rated significantly less similar to the reference audio for XTTS and LinaSpeech. No significant differences were observed for ZONOS between GE and N-GE similarity. The greatest effect sizes were seen when GE voices showed worse performance (LinaSpeech, d=.704; XTTS, d=.419), in comparison to when N-GE showed worse performance (F5TTS, d=-.311; E2TTS, d=-.283; CosyVoice2, d=-.226). A graph of these results and their distribution is also presented in Figure~\ref{fig:mushra_data}.

\begin{table}[h]
\centering
\captionsetup{justification=centering,singlelinecheck=false}
\resizebox{\columnwidth}{!}{%

\begin{tabular}{lcccc}
\hline
\multirow{2}{*}{Model} & \multicolumn{4}{l}{Human Ratings (Range 0-100)}  \\ 
\cline{2-5}
                       & GE    & N-GE    & t& d                         \\ 
\hline
F5TTS & 74.70 (19.58) & 67.80 (24.45) & -3.01$^\square$  & -.31\\
E2TTS & 72.13 (22.39) & 65.19 (26.42) & -2.74$^\square$  & -.28\\
CosyVoice2 & 62.57 (27.68) & 56.06 (29.81) & -2.19*  & -.23\\
ZONOS & 46.78 (28.69) & 50.07 (28.95) & 1.11  & .11\\
XTTS & 42.85 (30.29) & 55.01 (27.71) & 4.07$^\triangle$  & .42\\
LinaSpeech & 39.92 (27.94) & 58.90 (26.00) & 6.82$^\triangle$ & .70\\

\hline
\end{tabular}}
\caption{Analysis of Human Ratings, *=p\textless.05,$^\square$=p\textless.005,$^\triangle$=p\textless.001}

 \vspace{-10pt}
\label{tab:mushra_results}

\end{table}

\begin{table*}
\centering
\resizebox{\textwidth}{!}{%

\begin{tabular}{l|cccc|cccc|cccc|cc}
\hline
\multirow{2}{*}{Model} & \multicolumn{4}{l}{ECAPA-TDNN (Range 0-1)}& \multicolumn{4}{l}{TitaNet-L (Range 0-1)}& \multicolumn{4}{l}{ReDimNet-M (Range 0-1)}&\multicolumn{2}{l}{ICC}\\ \cline{2-15} 
                       & GE    & N-GE    & t & d  & GE    & N-GE    & t & d & GE    & N-GE    & t & d & GE    & N-GE                                 \\ \hline
F5TTS & .759 (.061)& .712 (.059) & –8.35$^\triangle$  & -.789 & .763 (.065)& .757 (.050)& -1.01& -.095& .771 (.052)& .752 (.042)& -4.26$^\triangle$&-.402 & .739 & .508\\
E2TTS & .754 (.063) & .707 (.064) & -7.85$^\triangle$  &-.742 & .788 (.059)& .763 (.055)& -4.54$^\triangle$& -.429& .771 (.050)& .740 (.048)& -6.70$^\triangle$&-.633 & .671 & .509\\
CosyVoice2 & .755 (.069) & .716 (.061) & -6.31$^\triangle$  & -.596 & .774 (.070)& .765 (.059)& -1.70& -.161& .761 (.050)& .735 (.042)& -5.87$^\triangle$&-.555 & .743 & .508\\
ZONOS & .644 (.077) & .624 (.055) & –2.83$^\square$  & -.267 & .680 (.097)& .698 (.055)& 2.43$^\square$& .229& .659 (.065)& .669 (.046)& 1.91& .180 & .703 & .365\\
XTTS & .536 (.104) & .529 (.104) & -.720  & -.068 & .636 (.069)& .635 (.069)& -.10& -.010& .609 (.069)& .596 (.063)& -2.03*& -.192 & .282 & .397\\
LinaSpeech & .384 (.115) & .457 (.100) & 7.15$^\triangle$ & .676 & .428 (.116)& .533 (.089)& 10.77$^\triangle$& 1.019& .392 (.097)& .472 (.072)& 9.74$^\triangle$& .922 & .778 & .584\\

\hline
\end{tabular}}
\caption{Analysis of speaker similarity, *=p\textless.05,$\square$=p\textless.005,$\triangle$=p\textless.001}

\label{tab:automated_results}
 \vspace{-20pt}
\end{table*}

\subsection{Speaker Similarity}

Speaker similarity from ECAPA-TDNN, TitaNet-L and ReDimnet-M are summarized in Table~\ref{tab:automated_results}. LinaSpeech consistently shows overall worse speaker similarity, with GE voices performing significantly worse than N-GE with moderate or high effect sizes for all embedding models. For XTTS, a significant difference is seen between GE and N-GE voices only for the ReDimNet-M model, where GE voices are predicted to have improved performance, though with a small effect size. Results for the ZONOS model had conflicting results for each embedding model, with ECAPA-TDNN predicting significantly better GE performance, TitaNet-L predicting significantly worse GE performance, and ReDimNet-M showing no significant differences. Conflicting results are similarly seen for F5TTS and CosyVoice2, where ECAPA-TDNN and ReDimNet-M had significantly increased speaker similarity for GE voices, but no difference was found for TitaNet-L. E2TTS consistently showed increased performance for GE voices across all models. ICC metrics found that agreement between similarity metrics was higher for GE voices than N-GE, except for XTTS, where N-GE voices had higher agreement.

\subsection{Automated MOS}

AMOS scores are presented in Table~\ref{tab:automated_mos_results}. For FT-Wav2Vec2, significant differences were found between the predicted scores for CosyVoice2, and LinaSpeech, with GE voices being predicted to have improved performance compared to N-GE voices. For UTMOS however, N-GE voices were predicted to have significantly greater AMOS scores for all models, with small to moderate effect sizes for all of these. ICC was also higher here for GE voices than for N-GE for all models. Further, for F5TTS and CosyVoice2, agreement was low for both datasets.

\begin{table*}
\centering

\begin{tabular}{l|cccc|cccc|cc}
\hline
\multirow{2}{*}{Model} & \multicolumn{4}{l}{FT-Wav2Vec2 (Range 1-5)}& \multicolumn{4}{l}{UTMOS (Range 1-5)}&\multicolumn{2}{l}{ICC}\\ \cline{2-11} 
                       & GE    & N-GE    & t & d  & GE    & N-GE    & t & d  & GE    & N-GE                                \\ \hline
F5TTS & 4.576 (.242) & 4.563 (.170) & .646  & -.061 & 4.192 (.260)& 4.284 (.201)& 4.202$^\triangle$& .398 &.237 &.125 \\
E2TTS & 4.236 (.426) & 4.229 (.296) & -.206  & -.019 & 3.978 (.331)& 4.065 (.280)& 2.993$^\square$& .283 &.548 &.421\\
CosyVoice2 & 4.691(.199) & 4.586(.186) & -5.719$^\triangle$  & -.542 & 4.366 (.151)& 4.405 (.120)& 3.026$^\square$& .316 & .261 &.244\\
ZONOS & 4.253 (.476) & 4.196 (.349) & 1.451  & .137 & 4.081 (.281)& 4.143 (.211)& 2.646*& .251 &.616 &.520\\
XTTS & 4.466 (.483) & 4.497 (.276) & .850  & .081 & 4.052 (.382)& 4.155 (.259)& 3.340$^\square$& .287 & .533&.283\\
LinaSpeech & 4.484 (.310) & 4.415 (.260) & -2.565* &  -.243 & 4.206 (.234)& 4.265 (.161)& 3.124$^\square$& .296 & .521 & .410\\

\hline
\end{tabular}
\caption{Analysis of AMOS, *=p\textless.05,$^\square$=p\textless.005,$^\triangle$=p\textless.001}

\label{tab:automated_mos_results}
 \vspace{-10pt}
\end{table*}

\subsection{Word Error Rate}

WER differences between GE and N-GE synthesized samples are summarized in Table ~\ref{tab:wer_results}. Significant differences were observed only for LinaSpeech, with WER being significantly higher for GE outputs. For F5TTS, outputs were near significant at p=.051, with similarly higher WER for GE than N-GE voices.

\begin{table}[H]
\centering
\captionsetup{justification=centering,singlelinecheck=false}

\begin{tabular}{lcccc}
\hline
\multirow{2}{*}{Model} & \multicolumn{4}{l}{WER  (Range 0-1)} \\ \cline{2-5}
                       & GE    & N-GE    & t & d                                \\ \hline
F5TTS      & .038 (.106)      & .022 (.059)        & –1.957               & -.185\\
E2TTS      & .040 (.121)      & .021 (.057)        & -2.167               & -.205\\
CosyVoice2 & .088 (.123)      & .087 (.131)        & -.028               & -.003\\
ZONOS      & .052 (.087)      & .060 (.098)        & .917                 & .087\\
XTTS       & .039 (.187)      & .020 (.077)        & -1.389               & -.132\\
LinaSpeech & .141 (.263)      & .072 (.112)        & -3.600$^\triangle$            & -.342\\ \hline
\end{tabular}%
\caption{Analysis of Word Error Rate, *=p\textless.05,$^\square$=p\textless.005,$^\triangle$=p\textless.001}

\label{tab:wer_results}
 \vspace{-10pt}
\end{table}

\section{Discussion}
Human evaluations using a MUSHRA-like evaluation found performance differences in speaker similarity of outputs between GE and N-GE voices which varied across models. For the models LinaSpeech and XTTS, participants rated synthesized GE voices as significantly worse recreations of the reference audio than N-GE voices, whereas for F5TTS, E2TTS, and CosyVoice2, the opposite results were observed, with N-GE voices being rated as significantly less similar to the reference audio. 


For automated metrics, these models disagree on the performance of the datasets depending on which evaluation model is used. Speaker similarity metrics had disparities in how they rated each dataset per model, with an overall significant difference between datasets for some speaker verification models which was not present for others. Most notably, all three speaker similarity metrics for the ZONOS model disagree on dataset performance, with ECAPA-TDNN predicting improved performance for GE voices, TitaNet-L predicting improved performance for N-GE voices, and ReDimNet-M finding no difference between either dataset. Automated MOS scores also show a similar divergence. The FT-Wav2Vec2 model only finds significant differences between the datasets for CosyVoice2 and LinaSpeech, where it predicts higher scores for GE voices, whereas the UTMOS model predicts significantly better performance for N-GE voices for all models. 

These disparities both between evaluation models and their differences with human ratings of speaker similarity draw into question the applicability of these as automated testing measures. These discrepancies likely stem from the training data used to develop these automated metrics, which may underrepresent queer and transgender voices, leading to difficulties in predicting actual performance. Consequently, although automated tools can detect differences, they do not appear to accurately reflect the qualitative judgments of human listeners. Without addressing these issues with automated metrics, it becomes difficult to assess a models actual performance on minority voice groups.

 We suggest a potential cause of the decreased performance could be that models which performed worse on subjective evaluations were trained on datasets containing read speech (XTTS, LinaSpeech), whereas two of the models (F5TTS, E2TTS) that showed better performance were trained on spontaneous speech datasets (no information is provided for other models). Spontaneous speech captures nuances of real speech in comparison to read speech, leading to a greater data diversity, even if the dataset doesn't explicitly contain gender-expansive speakers. Although the datasets used by these models do not state the queer-identities of their speakers, F5TTS and E2TTS were trained on the EMILIA dataset \cite{he2024emilia} which utilizes \say{in-the-wild} data, and so these datasets likely contain more diverse individuals. Poor replication of queer voices also happens with other TTS models \cite{campweshould}, in that `queer' aspects of these voices are removed from synthesized voices when suitable datasets are not used, which may come as a result of the model not being able to learn these aspects. This could also provide an explanation as to why agreement between automated metrics appears to be higher for GE voices for most TTS models, if analysis of speech samples follows a similar pattern in removing queer aspects of voices, and as such these speakers may end up clustering in similar score ranges between models, though this would require future research to investigate.



\subsection{Data Scarcity and Ethical Dilemmas in Curating Inclusive Datasets}
A potential cause of the issues presented here is due to the limited availability of queer speech datasets. We were only able to find two speech datasets explicitly containing queer data \cite{netzorg2024speech, hope_2023_midatlantic}, and only a single dataset that was suitable for assessing a zero-shot TTS model's performance on multiple queer voices, and even then, only 14 voices were available within the dataset utilized \cite{hope_2023_midatlantic}. Without large high quality datasets it is hard to expect that the models would be able to perform better, especially given that TTS models may fail to synthesize voices accurately  which are from sparser sections of a models latent space \cite{francis2025interpolating}.

A naive solution is to create a large dataset of these voices. However, this leads to concerns for queer communities. If larger, high quality, and well labeled datasets are made publicly available, the potential for abuse rises \cite{campweshould}. It is imperative to be aware and ensure that transgender and queer communities are actively engaged within the dataset creation process, as disclosure of identity of individuals within could cause harm to communities and individuals.
Moreover, strict guidelines surrounding data usage, privacy, and consent will be critical to ensure that risks associated with misuse are avoided. This could also be enhanced through anonymization techniques, controlled access protocols, or secure data repositories. Further, it is imperative that the principle of 'Nothing about us without us' is followed, meaning that those the dataset is supposed to represent must also be key contributors throughout the entire process. While the pursuit of more inclusive datasets may lead to significant technical gains, it's important to ensure that the needs and privacy of marginalized groups is placed at the forefront to avoid unintended harm. 

\section{Conclusion}
We investigated the ability of multiple zero-shot TTS models to synthesize speech of GE speakers in comparison to N-GE speakers, assessed via both human ratings and automated metrics. Further, we assessed whether automated metrics of speaker similarity and AMOS are effective for evaluating performance on GE speakers in comparison to subjective metrics, as well as how different evaluation model's consistency varies between GE and N-GE datasets. Subjective metrics varied across models, with two models showing worse performance for GE speakers, and three models showing better performance. The effect size was greater when GE voices performed worse. Various speaker similarity and AMOS models had performance predictions inconsistent with human scores, and themselves, where there is a disparity between the model's predictions on whether GE or N-GE voices perform better. These disparities could lead to assumptions that assessed models have better or worse performance on diverse voices than in reality. These results highlight a need to enhance both model and evaluation performance, as it is likely that the training data for these models underrepresents diverse voices. We argue for the creation of more inclusive datasets, however creation of these should be done with caution, keeping queer and transgender communities' needs at the forefront.

\section{Acknowledgements}
This work is funded by the WASP funded project Real-time
context-aware speech prosthesis for conversational interaction.

\section{Use of Generative AI Disclosure}
AI tools were used for minor revision and correction of portions of both text in this paper and code used for analysis, which authors thoroughly checked.

\bibliographystyle{IEEEtran}
\bibliography{mybib}

\end{document}